\documentclass[aps,prl,onecolumn,12pt,showpacs,superscriptaddress,amsmath,amssymb,nobibnotes,letterpaper]{revtex4-2}

\usepackage{blindtext}  
\usepackage{amssymb}
\usepackage{mathtools}
\usepackage[pdftex]{graphicx}
\usepackage[usenames, dvipsnames, svgnames, table]{xcolor}
\usepackage[colorlinks=true, citecolor=Blue,
            linkcolor=BrickRed, urlcolor=ForestGreen]{hyperref}
\usepackage{txfonts}
\usepackage{mathrsfs}
\usepackage{bm}
\usepackage{multirow}
\usepackage{color}
\usepackage{comment}
\usepackage{soul}

\usepackage{algorithm}  
\usepackage{algpseudocode}
\usepackage{CJK}

\begin{document}
\newcommand{\Com}[1]{{\color{red}{#1}\normalcolor}} 


\title{The Transformation-Response Framework: An Operational Reformulation of Quantum Mechanics}
\author{Meng-Jun Hu}
\email{humj@baqis.ac.cn}
\affiliation{Beijing Academy of Quantum Information Sciences, Beijing 100193, China}

\begin{abstract}

We present the transformation-response framework, an operational reformulation of quantum mechanics in which a quantum state is defined not as a mathematical object residing in an abstract Hilbert space, but as the catalog of a system's responses to all physical transformation operations. Concretely, for any operation $g$ drawn from the local group $G$ of the system, an interference experiment yields a complex value $\chi(g)$, which encodes how much the system overlaps with its original self after undergoing $g$. The complete collection of such responses $\{\chi(g): g\in G \}$, recorded for all possible operations, constitutes a function $\chi:G\rightarrow C$, which we term the characteristic function and identify as the quantum state itself.
The sole condition required for this catalog to represent a physical state is the positive-definiteness of $\chi(g)$, a condition that encodes the elementary operational requirement that no superposition of physical transformations may yield a negative probability. This single postulate is the only substantive assumption of the framework. From it, the entire mathematical apparatus of standard quantum mechanics is rigorously derived: the Hilbert space emerges via the Gelfand-Naimark-Segal construction, the Born rule follows from Bochner theorem, the Schr\"odinger equation arises from group automorphisms, and especially the Feynman path integral emerges as the Trotter continuum limit of the characteristic function. The framework is inherently background-independent and time-neutral, in which time appears not as an external parameter but as a coordinate along a specific one-parameter subgroup of $G$, thereby providing a natural language for quantum gravity. Crucially, because the framework is built on group structure, it yields a previously unrecognized physical constraint: product order positivity, which requires the characteristic function to remain positive-definite when restricted to any subset encoding a restricted operational sequence. This condition may lead to new, testable predictions. The transformation-response framework thus provides a conceptually unified, logically economical, and experimentally falsifiable foundation for quantum theory, rooted directly in operational primitives.

\end{abstract}

\maketitle

\section{1. Introduction} 
\label{sec1}
Standard quantum mechanics, as formalized by von Neumann in 1932 \cite{von}, rests upon a set of logically independent postulates: a quantum state is a ray in a Hilbert space, observables are self-adjoint operators, the Born rule assigns measurement probabilities, and the Schr\"odinger equation governs time evolution \cite{Dirac}.
This axiomatic structure has enjoyed extraordinary empirical success. From the spectroscopy of hydrogen to the precision tests of quantum electrodynamics, from the violation of Bell inequalities to the discovery of the Higgs boson, no experiment has exposed a flaw in the quantum formalism itself. Yet, beneath this triumphant phenomenological surface, the logical foundations of the theory harbor persistent conceptual tensions.

The first tension is logical redundancy. The postulates are independent of one another, the Born rule cannot be derived from the Schr\"odinger equation, nor can the Hilbert space structure be deduced from the other axioms. That a physical theory should require several separate foundational principles to describe a single class of phenomena suggests, by the usual standards of theoretical economy, that a deeper unifying principle remains undiscovered.
The second tension concerns the origin of probability. The Born rule is stated as a postulate, but why the probability should be the modulus squared of a complex amplitude, rather than any other function, is not explained. The theory can only answer that the rule works, not why it must be so.
The third and most urgent tension is the theory's complete reliance on a fixed, classical spacetime background. The Schr\"odinger equation depends on an external time parameter. The wavefunction 
$\Psi(x, t)$ is defined on a pre-existing spacetime manifold. When spacetime itself becomes a dynamical entity subject to quantization, this scaffolding collapses. The Wheeler-DeWitt equation of quantum cosmology describes a universe without an external time, and standard quantum mechanics offers no clear prescription for extracting physical predictions from such a timeless state \cite{Dewitt, Hawking, Rovel}. This is the problem of time, and it remains one of the most formidable obstacles to a coherent theory of quantum gravity.
These three tensions-logical redundancy, unexplained probability, and background dependence-point toward a common need: a reformulation of quantum mechanics that is logically economical, operationally grounded, and independent of any pre-existing spacetime stage. Attempts to reduce the number of postulates have been made, but these formulations remained mathematical reformulations without operational content and did not yield new physical constraints \cite{Feynman, Mackey, Ludwig}. The missing ingredient, a genuine operational foundation, has only become available with the rise of quantum information science \cite{Hardy, Chi}.

Over the past two decades, the rise of quantum information science has profoundly reshaped our understanding of what a quantum theory can be built upon. In quantum computation, the fundamental building blocks are not states but unitary gates, i.e., transformations applied to qubits, and an entire algorithm is specified by the sequence of operations performed, with the state playing only an auxiliary role. This way of thinking, designing and analyzing quantum processes entirely in terms of the operations that compose them, has become second nature to a generation of quantum information theorists. The discovery of indefinite causal order has reinforced this operational perspective from a different direction. Experiments with the quantum switch have demonstrated that the causal order of operations can itself be placed in a quantum superposition, a phenomenon that has no natural description in a formalism that presupposes a fixed background time. These experiments do not merely probe the limits of quantum mechanics; they invite us to reconsider what the fundamental building blocks of the theory ought to be. Quantum reference frame theory has contributed a complementary insight: when reference frames are themselves quantum systems, the very notion of a state becomes relational \cite{Rovelli}. There is no absolute state, only a state relative to a chosen frame. Taken together, these developments suggest that quantum mechanics, at its deepest level, is a theory of transformations and their relations, and that a formulation which takes operations,rather than states, as the primitive notion may capture this insight more faithfully than the conventional Hilbert space axiomatization. What has been lacking is a systematic reformulation of quantum mechanics that places this operational insight at the very foundation of the theory.

The transformation-response framework (TRF) presented here provides precisely such a reformulation. {\bf Its guiding physical philosophy is simple: we can only know a physical system by the operations we perform upon it and the responses we register}. We cannot inspect a quantum state directly; we can only apply transformation operations and observe the relative change of the physical system before and after, i.e., the responses to the operations. The response to a specific operation is mathematically described by a complex value that quantifies the overlap of the transformed and original system. These complex values constitute the raw operational data from which the entire quantum formalism should be constructed. 

{\bf Physically, the quantum state of a system is completely described by the catalog of system's responses to all possible physical transformation operations}. For each operation $g$, a complex response $\chi(g)$ of the system can be observed, which encodes how much the system overlaps with its original self after undergoing $g$. The sole condition required for this catalog of responses to represent a physically admissible quantum state is the mathematical property of positive-definiteness.
{\bf Mathematically, we thus define the quantum state of a system as a positive-definite function $\chi: G\rightarrow C$ on the locally compact group 
$G$ of physically realizable transformation operations}. We also refer to  $\chi(g)$ as the characteristic function. From this single postulate, we demonstrate that the entire apparatus of standard quantum mechanics can be rigorously derived: the Hilbert space via the  Gelfand–Naimark–Segal (GNS) construction, the Born rule via Bochner's theorem, the Schr\"odinger equation via group automorphisms, and the Feynman path integral as the Trotter limit of the characteristic function.

The TRF is inherently background-independent and time-neutral, which makes it naturally suited for quantum gravity, where the absence of an external time parameter is a central conceptual challenge. Crucially, because the framework is built upon group structure, it gives rise to a previously unrecognized physical constraint, product order positivity, which requires that the characteristic function remain positive-definite when restricted to any subset encoding a restricted order of operations. By encoding operational restrictions, such as fixing a causal order or limiting the order of admissible operations, directly into the algebraic structure of the framework, the product order positivity may provide a new theoretical language for formulating physical constraints that are invisible to the traditional framework.

\section{2. The Operational Origin of Quantum States: Motivation and Postulate} 
\label{sec2}
In this section, we formulate the physical motivation for the TRF and state its core postulate. The guiding principle is that a quantum state is not an entity that we inspect directly. It is something defined through the operations we perform and the responses we register. Our task is to turn this operational intuition into a precise mathematical condition, without presupposing any of the structures, Hilbert space, state vectors, Born rule, that we intend to derive.

\subsection{2.1. From Operations to States: The Response Catalog}

The experimenter never ``sees" a quantum state directly. What the experimenter does is take a system from a standardized physical source, apply controlled physical transformations to it, and register measurement outcomes. This sequence—source, transformation, outcome—is the raw material of all experimental physics. The operationalist doctrine, tracing back to Bridgman \cite{Bridgman} and central to modern quantum information theory, holds that a physical concept means nothing more than the set of operations used to measure it \cite{Nielsen}. Applied to quantum mechanics, this demands a reordering of logical priorities: instead of beginning with an abstract Hilbert space and defining operations as secondary actions within it, we begin with the concrete operations themselves and construct the Hilbert space as a derived structure.

What is the most elementary operational datum we can collect about a quantum system?
Consider an experimental arrangement in which a probe, a measurement apparatus with a pointer degree of freedom, is coupled to the system. The probe is prepared in a fixed reference configuration, and the probe-system interaction is engineered so that the evolution of the system depends on the state of the pointer. Specifically, the coupling is designed so that when the pointer is in a designated reference configuration, the system undergoes no transformation; when the pointer is displaced from this reference, the system undergoes a specific transformation operation $g$, drawn from the set $G$ of all physical transformations that can be applied to the system. By preparing the pointer in a coherent superposition of these two configurations and subsequently measuring an appropriate pointer observable, we obtain an interference signal. The displacement of the interference fringes relative to the reference configuration yields one real number; the visibility of the fringes yields another. Together, these two numbers constitute a complex number. We denote this complex number by $\chi(g)$ and call it the response of the system to the transformation $g$.

Several features of this procedure are essential. First, no notion of a quantum state has been invoked. The entire description refers only to the preparation of the probe, the controlled probe-system coupling, and the readout of the pointer. The complex number $\chi(g)$ is a directly measurable operational quantity, extracted from pointer statistics. Second, the procedure is platform-independent. In the circuit model of quantum computation, it reduces to the Hadamard test, where an ancillary qubit controls the application of $g$ and the real and imaginary parts of $\chi(g)$ are obtained from Pauli expectation values. In quantum optics, it corresponds to phase-sensitive probe-system interferometry. In trapped ions, it is realized via Ramsey interferometry with a control ion. The operational blueprint is universal. Third, the set 
$G$ of all physical transformations naturally forms a group. Two transformations can be performed in sequence, yielding a third; the identity transformation leaves the system unchanged; every transformation can be reversed. These properties, closure, associativity, identity, and invertibility are empirical facts about how physical operations compose in the laboratory, not mathematical abstractions imposed upon the physics.

We now perform this procedure for every transformation $g\in G$. For each $g$, we record the complex response $\chi(g)$. The result is a function $\chi: G\rightarrow C$, which we term the characteristic function of the system. This function is the complete catalog of the system's responses to all possible physical transformations.
In the operational framework, there is no hidden ``state itself" lurking behind this catalog. Every statement we can make about the system, every prediction, every explanation, every comparison with theory, must be expressible in terms of the responses $\{\chi(g) \}$. The catalog exhausts the operational content of the system. We therefore make the central definition of the TRF:

{\bf Definition of State in Transformation-Response Framework}: The quantum state of a system is described by the complete catalog of its responses to all possible physical transformations. Mathematically, a quantum state is a normalized positive-definite function $\chi: G\rightarrow C$.

This definition inverts the traditional logical order. In standard quantum mechanics, one begins with a Hilbert space $\mathcal{H}$, a state vector $|\psi\rangle\in\mathcal{H}$, and a unitary representation $U$ of $G$ on $\mathcal{H}$, and then derives the characteristic function as $\chi(g)=\langle\psi|U(g)|\psi\rangle$. In the TRF, $\chi(g)$ is the primary object. The Hilbert space, the state vector, and the unitary representation will be constructed from $\chi$. This operational definition of a state is universal. In classical physics, we never ``see'' a state directly either, we infer it from how the system responds to operations. The distinction between classical and quantum lies not in what a state is, but in the structure of the group $G$ that defines it. When 
$G$ is abelian, the positive-definiteness condition forces the state space into a simplex-a classical probability space whose extreme points are definite configurations, admitting no superpositions. When 
$G$ is non-abelian, the non-commutativity of operations breaks the simplex structure \cite{Birkhoff}. Superposition, interference, and entanglement all flow from this single algebraic fact.

\subsection{2.2. The Sole Condition: Positive-Definiteness as Physical Consistency}

Not every function $\chi: G\rightarrow C$ can serve as a physically admissible quantum state. The response catalog must satisfy a consistency condition imposed by the physics of superposition.

Consider a more sophisticated experiment. Instead of superposing ``do nothing" and ``perform transformation $g$" on the probe, we superpose a finite collection of transformations. We prepare the probe in a superposition of configurations, each corresponding to the application of a different transformation $g_{i}$ to the system, with complex amplitudes $c_{i}$. After the controlled interaction, we measure the probe in a basis that asks whether the system has returned to its reference configuration. {\bf The operational requirement is elementary: no superposition of physical transformations may yield a negative probability}. The probability that the system, after undergoing a coherent superposition of transformations, is found to have returned to its reference configuration must be non-negative.
For any finite set of transformations $\{ g_{1}, g_{2},..., g_{n}\}\subset G$ and any complex coefficients $\{c_{1}, c_{2},..., c_{n}\}$, this requirement translates directly into the mathematical inequality
\begin{equation}
\sum_{i=1}^{n}\sum_{j=1}^{n}c_{i}^{*}c_{j}\chi(g_{i}^{-1}g_{j})\ge 0
\end{equation}
This is the defining condition for 
$\chi$ to be a positive-definite function on the group $G$. If we define a matrix $\mathbf{M}$ with element $\mathbf{M}_{ij}=\chi(g_{i}^{-1}g_{j})$, then equation (1) becomes $\vec{c}^{\dagger}\mathbf{M}\vec{c}\ge 0$ with $\mathbf{M}^{\dagger}=\mathbf{M}$. It is not a mathematical artifice, but rather the direct encoding of the physical principle that probabilities are non-negative, which applies not merely to individual measurement outcomes, but to the coherent superpositions of operations that quantum world uniquely permits.
The positive-definiteness condition (1) is the sole substantive postulate of the TRF. No further assumptions about the nature of the system are required.

{\bf Core Postulate of Transformation-Response Framework}:
Let $G$ be the group of all physical transformations that can be applied to a system. A quantum state of the system is described by the function $\chi:G\rightarrow C$ satisfying:

{\bf (1). Normalization}: $\chi(e)=1$, where $e\in G$ is the identity transformation. The response to ``doing nothing" is certainty that the system remains in its reference configuration.

{\bf (2). Positive-definiteness}: For every finite set $\{ g_{1}, g_{2},..., g_{n}\}\subset G$ and every set of complex coefficients $\{c_{1}, c_{2},..., c_{n}\}$, $\sum_{i=1}^{n}\sum_{j=1}^{n}c_{i}^{*}c_{j}\chi(g_{i}^{-1}g_{j})\ge 0$.
 
{\bf (3). Continuity}: $\chi$ is continuous with respect to the natural topology on $G$. Operationally, two transformations that are arbitrarily close must yield arbitrarily close responses. The continuity encodes the operational requirement that transformations indistinguishable by any finite-precision experiment must elicit indistinguishable responses.

This is the entire foundational structure of the TRF. Traditional quantum mechanics requires several independent postulates. However, all of them can emerge as theorems from the single postulate stated in the TRF, as we demonstrate in the subsequent sections.

\subsection{2.3. A First Example: The $Z_{2}$ System}

To give the abstract postulate a concrete anchor, we examine the simplest non-trivial group: $G=Z_{2}=\{e, g\}$, where $g^{2}=e$. This describes a system on which only one non-trivial reversible transformation can be performed. For instance, a two-level system on which we can only apply a $\pi$-pulse around a fixed axis.
According to our definition, the quantum state is described by the function $\chi:\{e, g\}\rightarrow C$ with $\chi(e)=1$. The positive-definiteness condition must hold for all finite subsets. The only non-trivial constraint arises from the subset $\{e, g\}$ is
$|c_{1}|^{2}+|c_{2}|^{2}+c_{1}^{*}c_{2}\chi(g)+c_{2}^{*}c_{1}\chi(g^{-1})\ge 0$ with arbitrary coefficients $c_{1}, c_{2}\in C$. This is equivalent to require a Hermitian matrix $\mathbf{M}$ to be positive semi-definite with:
\begin{equation}
\mathbf{M}=\begin{bmatrix}
1 & \chi(g) \\
\chi(g^{-1}) & 1
\end{bmatrix}.
\end{equation}
The positive semi-definite requires $\mathrm{det} \mathbf{M}=1-\chi(g)\chi(g^{-1})=1-|\chi(g)|^{2}\ge 0$, hence $|\chi(g)|\le 1$.

The set of all quantum states of a $Z_{2}$ system is therefore the closed unit disk in the complex plane. This set is convex and the extreme points lie on the boundary circle $|\chi(g)|=1$ is pure states. Every point in the interior $|\chi(g)|< 1$ is a convex combination of boundary points and thus corresponds to a mixed state. The centre of the disk, $|\chi(g)|=0$, is the unique maximally mixed state: the response to the non-trivial operation $g$ vanishes entirely, signifying that the system retains no information about whether $g$ was applied.

\section{3. Reconstructing the Hilbert Space: The GNS Construction} 
\label{sec3}

Now we demonstrate how the entire Hilbert space structure of quantum mechanics emerges from the TRF postulate alone. The mathematical engine is the Gelfand–Naimark–Segal construction \cite{GNS}. The physical result is that every admissible response catalog $\chi$ automatically generates a Hilbert space $\mathcal{H}_{\chi}$, a unitary representation $U_{\chi}$ of $G$ on $\mathcal{H}_{\chi}$ , and a distinguished cyclic vector $|\Omega\rangle_{\chi}$ from which $\chi(g)$ is recovered as $\chi(g)=_{\chi}\langle\Omega|U_{\chi}(g)|\Omega\rangle_{\chi}$. 

The first step is to construct a group algebra $C[G]$, which can be done by considering all formal finite linear combinations of group elements with complex coefficients:
\begin{equation}
\mathcal{A}=\{\sum_{i=1}^{n}c_{i}g_{i} | n\in N, c_{i}\in C, g_{i}\in G\}.
\end{equation}
Physically, an element $\sum_{i}c_{i}g_{i}\in\mathcal{A}$ represents a coherent superposition of physical transformations. $\mathcal{A}$ is a complex vector space under the obvious operations of addition and multiplication:
\begin{equation}
\begin{aligned}
& \sum_{i} a_{i}g_{i}+\sum_{i} b_{i}g_{i}=\sum_{i}(a_{i}+b_{i})g_{i} \\
& (\sum_{i} a_{i}g_{i})\cdot (\sum_{j} b_{j}h_{j}) = \sum_{i,j}a_{i}b_{j}(g_{i}h_{j}).
\end{aligned}
\end{equation}

The second step is to define inner product \cite{Stine}. The response catalog $\chi$ provides a natural candidate for an inner product on $\mathcal{A}$. For two superpositions $A=\sum_{i}a_{i}g_{i}$ and $B=\sum_{j}b_{j}h_{j}$ in $\mathcal{A}$, we define
\begin{equation}
\langle A, B\rangle_{\chi} = \sum_{i,j}a_{i}^{*}b_{j}\chi(g_{i})^{*}\chi(h_{j})=\sum_{i,j}a_{i}^{*}b_{j}\chi(g_{i}^{-1}h_{j}),
\end{equation}
where $\chi(g)^{*}=\chi(g^{-1})$ is used. For $A=B$, we recover the positive-definiteness $\sum_{i,j}a_{i}^{*}a_{j}\chi(g_{i}^{-1}g_{j})\ge 0$. Thus, the positive-definiteness postulate guarantees that $\langle\cdot,\cdot\rangle_{\chi} $ is a genuine pre-inner product on $\mathcal{A}$. A pre-inner product may admit non-zero vectors with vanishing norm. These null vectors constitute a subspace $\mathcal{N}_{\chi}=\{a\in\mathcal{A}|\langle a , a\rangle_{\chi}=0\}$. Physically, a superposition $a$ with $\langle a , a\rangle_{\chi}=0$ is one for which the operation superposition produces no detectable response, which is operationally indistinguishable from the ``zero superposition". It is therefore physically natural to identify null vectors with zero.

The standard mathematical procedure is to form the quotient space $\mathcal{H}_{\chi}^{0}=\mathcal{A}/\mathcal{N}_{\chi}$, in which an element of $\mathcal{H}_{\chi}^{0}$ is an equivalence class $[A]=A+\mathcal{N}_{\chi}$. The pre-inner product descends to a well-defined, positive-definite inner product on the quotient  space $\mathcal{H}_{\chi}^{0}$:
\begin{equation}
\langle A, B\rangle_{\chi} = \langle[A],[B]\rangle_{\mathcal{H}_{\chi}^{0}},
\end{equation}
which is independent of the choice of representatives. This inner product is now strictly positive, $\langle[A],[A]\rangle_{\mathcal{H}_{\chi}^{0}}=0$ if and only if $[A]=0$. Finally, the GNS Hilbert space $\mathcal{H}_{\chi}$ is defined as the completion of $\mathcal{H}_{\chi}^{0}$ with respect to the norm induced by this inner product:
\begin{equation}
\mathcal{H}_{\chi}=\bar{\mathcal{H}_{\chi}^{0}}^{||\cdot||},  \qquad      ||\psi||=\sqrt{\langle\psi,\psi\rangle_{\mathcal{H}_{\chi}^{0}}}.
\end{equation}

We thus have constructed a genuine Hilbert space $\mathcal{H}_{\chi}$ from the single datum of a positive-definite function $\chi$ on $G$. The construction yields, by its very nature, two further essential structures. Consider the identity element $e\in G$, regarded as an element of $\mathcal{A}$. Let $|\Omega\rangle_{\chi}=[e]=e+\mathcal{N}_{\chi}\in\mathcal{H}_{\chi}$, we have $_{\chi}\langle\Omega|\Omega\rangle_{\chi}=\chi(e^{-1}e)=\chi(e)=1$, which is a unit vector. It is the mathematical representative of the system's initial reference configuration—the system prepared by the fixed physical source and the fixed pre-processing protocol, with no additional operation applied. The group $G$ acts naturally on $\mathcal{A}$ by left multiplication $g\cdot(\sum_{i}a_{i}g_{i})=\sum_{i}a_{i}(gg_{i})$, which preserves the algebraic structure and the inner product $\langle g\cdot A, g\cdot B\rangle_{\chi}=\langle A, B\rangle_{\chi}$. Moreover, it maps the null space $\mathcal{N}_{\chi}$ to itself. Consequently, the action descends to a well-defined linear operator $U_{\chi}(g)$ on the quotient $\mathcal{H}_{\chi}^{0}$: $U_{\chi}(g)[A]=[g\cdot A]$. Since the inner product is preserved, each $U_{\chi}(g)$ is an isometry. The group property $U_{\chi}(gh)=U_{\chi}(g)U_{\chi}(h)$ follows directly from the associativity of group multiplication. The inverse is $U_{\chi}(g)^{-1}=U_{\chi}(g^{-1})$ , so each $U_{\chi}(g)$ is unitary. Extending by continuity, we obtain a unitary representation of $G$ on the completed Hilbert space $\mathcal{H}_{\chi}$: $U_{\chi}: G\rightarrow \mathcal{U}(\mathcal{H}_{\chi})$. The vector $|\Omega\rangle_{\chi}$ is cyclic for the representation $U_{\chi}$. By construction, the set $\{U_{\chi}(g)|\Omega\rangle_{\chi}|g\in G\}=\{[g]|g\in G\}$ spans a dense subspace of $\mathcal{H}_{\chi}$. Physically, this means that every vector in the Hilbert space can be approximated arbitrarily well by applying some operation $g$ to the reference configuration and forming superpositions of the resulting states. The entire Hilbert space is thus generated by acting on the reference state with the group. The cyclic vector $|\Omega\rangle_{\chi}$ and the unitary representation $U_{\chi}$ reconstruct the response catalog in Hilbert space as:
\begin{equation}
\chi(g)=\chi(e^{-1}g)=\langle e, g\rangle_{\chi}=\langle[e], [g]\rangle_{\mathcal{H}_{\chi}}=_{\chi}\langle\Omega|U_{\chi}(g)|\Omega\rangle_{\chi}
\end{equation}
The response to a transformation $g$ is precisely the expectation value of the unitary operator $U_{\chi}(g)$ in the reference state.

\section{4. Probability without Postulates: The Born Rule from Bochner Theorem}
\label{sec5}
In standard quantum mechanics, the Born rule is an independent postulate: given a state vector $|\psi\rangle$ and a measurement associated with a self-adjoint operator $\hat{A}$ with spectral decomposition $\hat{A}=\sum_{k}a_{k}|a_{k}\rangle\langle a_{k}|$, the probability of obtaining outcome $a_{k}$ is $P(a_{k})=\langle\psi|a_{k}\rangle\langle a_{k}|\psi\rangle=|\langle a_{k}|\psi\rangle|^{2}$. Why the probability should be given by the modulus squared of a complex amplitude, rather than any other function, is left unexplained.
In the TRF, however, the Born rule is not a postulate, but rather derived from the positive-definiteness condition via Bochner theorem \cite{Bochner} and its non-commutative generalization. The derivation reveals that the Born rule is the unique probability assignment compatible with the requirement that no superposition of physical transformations may yield a negative probability.

In the TRF, measuring an observable corresponds to applying a specific family of transformation operations $g(s)$ with $s$ being a continuous real parameter and analyzing the resulting response catalog $\{\chi(g(s))\}$. Obviously, when an experimentalist apply operation $g(s_{1})$ first and then operation $g(s_{2})$ equivalent to operation $g(s_{1}+s_{2})$, which implies that the operations $\{g(s)\}$ form a one-parameter subgroup of $G$.  Since $g(s_{1}+s_{2})=g(s_{2})\circ g(s_{1})=g(s_{1})\circ g(s_{2})$, this subgroup is Abelian. The GNS construction then represents this operational subgroup $\{g(s)\}$ as a strongly continuous one-parameter unitary group $U_{\chi}(g(s))$ on the emergent Hilbert space $\mathcal{H}_{\chi}$. According to the Stone theorem \cite{Stone}, there exists a unique self-adjoint (Hermitian) operator $\hat{A}$ such that $U_{\chi}(g(s))=e^{-i\hat{A}s}$ in which $\hat{A}$ is actually the generator of group. Restrict the characteristic function $\chi$ to the Abelian subgroup generated by $\hat{A}$, we have $\chi_{A}(s)=\chi(U_{A}(g(s)))=\chi(e^{-i\hat{A}s})=_{\chi}\langle\Omega|e^{-i\hat{A}s}|\Omega\rangle_{\chi}$ with $\chi_{A}(g(0))=\chi_{A}(e)=1$. The $\chi_{A}(s)$ is thus a positive definite continuous function on $\mathbb{R}$ with $\chi_{A}(0)=1$.  The Bochner's theorem guarantees the existence of a unique finite positive Borel measure $\mu_{A}$ on $\mathbb{R}$ such that:
\begin{equation}
\chi_{A}(s)=\int_{\mathbb{R}}e^{-ias}d\mu_{A}(a),
\end{equation}
where the measure $\mu_{A}$ is a probability measure $\mu_{A}(\mathbb{R})=\chi_{A}(0)=1$. On the other hand, the self-adjoint operator $\hat{A}$ admits a spectral decomposition $\hat{A}=\int_{\mathbb{R}}adE(a)$ with $E(a)$ is a projection-valued measure on $\mathbb{R}$. For any Borel set $B\subset R$, the projection $E(B)$ represents the ``yes/no" measurement of whether the value of $A$ lies in $B$. In the GNS constructed Hilbert space $\mathcal{H}_{\chi}$, we thus have
\begin{equation}
\chi_{A}(s)=_{\chi}\langle\Omega|e^{-i\hat{A}s}|\Omega\rangle_{\chi}=_{\chi}\langle\Omega|\int_{\mathbb{R}}e^{-ias}dE(a)|\Omega\rangle_{\chi}=\int_{\mathbb{R}}e^{-ias}d_{\chi}\langle\Omega|E(a)|\Omega\rangle_{\chi},
\end{equation}
which implies probability measure $\mu_{A}(a)={\chi}\langle\Omega|E(a)|\Omega\rangle_{\chi}$. We have therefore derived the Born rule from the positive-definiteness postulate alone. The probability that a measurement of $A$ yields a value in a Borel set $B$ is not an independent axiom but a theorem: it equals the expectation value of the spectral projection $E(B)$ evaluated in the GNS state $|\Omega\rangle_{\chi}$. For a pure state, $|\Omega\rangle_{\chi}=|\psi\rangle$ is the unique (up to phase) state vector characterizing the system. If the dimension of $\hat{A}$ is finite and admits a discrete spectrum, then $E(a_{k})=\Pi_{k}=|a_{k}\rangle\langle a_{k}|$, the probability of obtain $a_{k}$ becomes $P(a_{k})=\mu_{A}(a_{k})=_{\chi}\langle\Omega|\Pi_{k}|\Omega\rangle_{\chi}=|\langle a_{k}|\psi\rangle|^{2}$. Thus, the modulus squared of the complex amplitude—the hallmark of the Born rule—is not an arbitrary choice. It is forced upon us by the conjunction of two mathematical facts: Bochner's theorem, which guarantees the existence of a probability measure from the positive-definite function $\chi_{A}$; and Stone's theorem together with the spectral theorem, which identifies this measure as the spectral measure of the corresponding self-adjoint operator evaluated in the GNS state.

A physical system described by a non-abelian group 
$G$ possesses many distinct one-parameter subgroups, each corresponding to a different observable—position, momentum, angular momentum, energy, and so on. These subgroups need not commute. Nevertheless, each is abelian internally, so the restriction of $\chi$ to any one of them yields, via Bochner theorem, a well-defined probability measure for that observable. The GNS construction provides a single Hilbert space $\mathcal{H}_{\chi}$ and a single reference state $|\Omega\rangle_{\chi}$ on which all these observables are simultaneously represented. The unitary representation $U_{\chi}$ maps each one-parameter subgroup $\{g(s)\}$ to a strongly continuous unitary group $U_{\chi}(g(s))=e^{-i\hat{A}s}$, and Stone theorem assigns a self-adjoint generator $\hat{A}$ to each. The spectral measure of $\hat{A}$ evaluated in $|\Omega\rangle_{\chi}$ gives the Born rule for $A$.

The consistency of this assignment, that the probability measures for different observables all derive from the same state $|\Omega\rangle_{\chi}$, is guaranteed by the fact that they are all restrictions of the same globally positive-definite function $\chi$ on the full group $G$. The positive-definiteness condition on $G$ is strictly stronger than positive-definiteness on each abelian subgroup separately. It forces the Stone generators of non-commuting one-parameter subgroups to obey the canonical commutation relations \cite{Wigner}, which are mathematically equivalent to the uncertainty principle. Any hypothetical joint probability distribution for non-commuting observables would correspond to a function $\chi$ that fails to be positive-definite on 
$G$, and therefore does not describe a physically admissible quantum state.

In standard quantum mechanics, observables are are postulated to be Hermitian operators. {\bf In the TRF, however, the primitive observable is not a Hermitian operator but a unitary operator $U_{\chi}$ itself under the GNS representation}. Hermitian observables emerge as a derived special case. When the experimentalist does not apply a single operation in isolation, but instead scans a continuous family of operations forming a one-parameter subgroup $\{g(s)\}\subset G$, Bochner theorem and Stone theorem together extract a self-adjoint generator $\hat{A}$ from the $s$-dependence of the corresponding unitaries $U(g(s))=e^{-i\hat{A}s}$. This generator is Hermitian, but its Hermiticity is a theorem, not an axiom. More general operational procedures also naturally yield positive operator-valued measures (POVMs), whose positive elements descend from positive elements of the group algebra under the GNS representation. In all cases, the sole constraint is the positive-definiteness of the state. The framework thus derives what an observable can be, rather than postulating it.

\section{5. Dynamics without Externel Time: The Sch\"odinger Equation as Group Automorphism}
\label{sec51}

In standard quantum mechanics, the Schrödinger equation 
$i\hbar d|\psi\rangle/dt=\hat{H}|\psi\rangle$ is postulated as an independent dynamical law. The time parameter $t$ and the Planck constant $\hbar$ both appear as fundamental ingredients whose origins are left unexplained. When spacetime itself becomes dynamical,as in quantum gravity, this external time parameter ceases to be available, and the Schr\"odinger equation in its standard form loses its foundation. 
In the TRF, we do not postulate the Schr\"odinger equation. Instead, we derive a {\bf general equation of motion} that holds for any continuous family of transformations, with no privileged time parameter and no pre-existing Planck constant. The Planck constant emerges naturally as the universal scale factor connecting abstract group parameters to laboratory coordinates. The Schrödinger equation then appears as merely one instance of this general equation, in which the transformation parameter is identified as the reading of a clock.

\subsection{5.1. Dynamics as Group Automorphisms}
A quantum state in the TRF is a positive-definite function $\chi:G\rightarrow C$. The only structure available to formulate dynamics is the group $G$ itself. There is no external time, no background spacetime, no pre-existing Hamiltonian. A change in the physical situation, for example a different experimental apparatus, a different external field, a different reference frame, corresponds to a restructuring of the set of available operations. Mathematically, such a restructuring is a group automorphism: a bijection $\alpha: G\rightarrow G$ that preserves the group structure:
\begin{equation}
\alpha(gh)=\alpha(g)\alpha(h), \qquad \alpha(e)=e.
\end{equation}
Under this change, the response catalog transforms by pullback:
\begin{equation}
\chi_{\alpha}(g)=\chi(\alpha(g)).
\end{equation}
This is the most general kinematical transformation of a state. No differential equation, no parameter, no constant has been introduced.

We now consider a continuous family of such automorphisms. Let $\{\alpha_{s}\}_{s\in\mathbb{R}}$ be a one-parameter family of automorphisms of $G$, labelled by a real parameter $s$, satisfying $\alpha_{s_{1}}\circ\alpha_{s_{2}}=\alpha_{s_{1}+s_{2}}$ and $\alpha_{0}=I_{G}$. The state evolves along this family as $\chi_{s}(g)=\chi_{0}(\alpha_{s}(g))$. For example, if the parameter $s$ is chosen as time $t$, the group automorphism corresponds to the time evolution which described by the group element, i.e., time translation operation $h_{t}$. The time-evolved state in the TRF is given by $\chi_{t}(g)=\chi_{0}(h_{t}^{-1}gh_{t})$ with $\alpha_{t}(g)=h_{t}^{-1}gh_{t}$. At the infinitesimal level, the automorphism is generated by a derivation $D$ acting on functions on $G$. For any smooth function $f: G\rightarrow C$, the derivative along the automorphism at the identity is defined as:
\begin{equation}
(Df)(g)=\frac{d}{ds}f(\alpha_{s}(g))|_{s=0},
\end{equation}
where $D$ satisfies the Leibniz rule $D(f_{1}f_{2})=(Df_{1})f_{2}+f_{1}(Df_{2})$. Applying this to the state $\chi(g)$, we obtain the general equation of motion on the group
\begin{equation}
\boxed{\frac{d}{ds}\chi_{s}(g)=D\chi_{s}(g)}.
\end{equation}
This is the {\bf fundamental dynamical law}. It involves no Hilbert space, no wavefunction, and no privileged parameter. The variable $s$ is simply the coordinate along the chosen family of automorphisms.

\subsection{5.2. The Hilbert Space Equation of Motion and the Emergence of Planck Constant}

To connect with the familiar language of quantum mechanics, we now apply the GNS construction. The state $\chi$ generates a Hilbert space $\mathcal{H}_{\chi}$, a reference vector $|\Omega\rangle_{\chi}$, and a unitary representation $U_{\chi}$ of $G$. The one-parameter family of automorphisms $\alpha_{s}$ is implemented by a one-parameter unitary group $V(s)$ on $\mathcal{H}_{\chi}$. The relationship between the automorphism and its unitary implementation is given by
\begin{equation}
U_{\chi}(\alpha_{s}(g))=V(s)^{\dagger}U_{\chi}(g)V(s),
\end{equation}
where $V(s_{1}+s_{2})=V(s_{1})V(s_{2})$. By the continuity postulate and Stone's theorem, there exists a self-adjoint generator $\hat{K}$ such that $V(s)=e^{-i\hat{K}s}$. Defining the evolved reference vector as
\begin{equation}
|\Omega_{s}\rangle=V(s)|\Omega\rangle_{\chi}=e^{-i\hat{K}s}|\Omega\rangle_{\chi},
\end{equation}
The evolved state $\chi_{s}(g)$ becomes
\begin{equation}
\chi_{s}(g)=\chi_{0}(\alpha_{s}(g))=_{\chi}\langle\Omega|U_{\chi}(\alpha_{s}(g))|\Omega\rangle_{\chi}=\langle\Omega_{s}|U_{\chi}(g)|\Omega_{s}\rangle.
\end{equation}
The vectors $|\Omega_{s}\rangle$ obviously satisfy a first-order differential equation:
\begin{equation}
i\frac{d}{ds}|\Omega_{s}\rangle=\hat{K}|\Omega_{s}\rangle.
\end{equation}
The dynamical equation of $\chi_{s}(g)$ in the Hilbert space $\mathcal{H}_{\chi}$, according to Eq. (17) and Eq. (18), can be derived as
\begin{equation}
\frac{d}{ds}\chi_{s}(g)=i\langle\Omega_{s}|[\hat{K}, U_{\chi}(g)]|\Omega_{s}\rangle.
\end{equation}
This equation shows that the rate of change of state $\chi_{s}(g)$ along the parameter $s$ is governed by the commutator of the generator $\hat{K}$ with the unitary operator $U_{\chi}(g)$. It is exactly equivalent to the group-level equation of motion given by Eq. (14), with the derivation $D$ realized in the GNS representation as the inner derivation $D\chi_{s}(g)=i\langle\Omega_{s}|[\hat{K}, U_{\chi}(g)]|\Omega_{s}\rangle$. This equivalence guarantees the strict correspondence between the abstract group dynamics and the concrete operator dynamics on the emergent Hilbert space.

{\bf The emergence of Planck constant}. To connect above abstract equation to laboratory measurements, the experimentalist must calibrate the parameter $s$ against a physical coordinate. For any one-parameter subgroup, a linear relation should be established between the abstract parameter and the physical quantity read from instruments. For time evolution, $s=wt$ with $w$ is a frequency and $t$ is the reading of clock; For spatial translations, $s=kx$ with $k$ is a wave number and $x$ is the displacement measured by ruler; For rotations, $s=\theta$ with $\theta$ is the angle measured by goniometer. Substituting these calibrations into the abstract equation of motion yields, for each case,
\begin{equation}
i\frac{d}{dt}|\Omega_{t}\rangle=w\hat{K}|\Omega_{t}\rangle, \quad i\frac{d}{dx}|\Omega_{x}\rangle=k\hat{K}|\Omega_{x}\rangle, \quad i\frac{d}{d\theta}|\Omega_{\theta}\rangle=\hat{K}|\Omega_{\theta}.
\end{equation}
The experimentalist now defines the physical observables by absorbing the calibration factor into the generator, introducing a universal conversion constant $\hbar$ with dimensions of action: $\hat{H}=\hbar w\hat{K}, \hat{P}=\hbar k\hat{K}, \hat{J}=\hbar\hat{K}$, The equations of motion then take the form
\begin{equation}
i\hbar\frac{d}{dt}|\Omega_{t}\rangle=\hat{H}|\Omega_{t}\rangle, \quad i\hbar\frac{d}{dx}|\Omega_{x}\rangle=\hat{P}|\Omega_{x}\rangle, \quad i\hbar\frac{d}{d\theta}|\Omega_{\theta}\rangle=\hat{J}|\Omega_{\theta}.
\end{equation}
The above equations are the general Hilbert space equation of motion with physical units and the first one is the familiar Schr\"odinger equation. The Planck constant $\hbar$ is the same for all three equations. This universality is not an additional postulate, it is enforced by the global positive-definiteness of $\chi$ on the full non-abelian group $G$. If different subgroups required different conversion constants, the characteristic function evaluated on mixed sequences of operations from distinct subgroups would fail the positive-definiteness condition. {\bf The constant $\hbar$ therefore emerges as the unique scale factor that makes the representation of the entire group $G$ consistent with a single positive-definite state}. The set of Eq. (21) reveals a profound unification. All continuous dynamics in quantum mechanics—time evolution, spatial translations, rotations, boosts, gauge transformations—are governed by the same mathematical structure:
\begin{equation}
i\hbar\frac{d}{d\lambda}|\Omega_{\lambda}\rangle=\hat{O}|\Omega_{\lambda}\rangle,
\end{equation}
where $\lambda$ is any physical coordinate along a one-parameter subgroup of $G$, and $\hat{O}$ is the corresponding physical observable. The special status of the Schr\"odinger equation is purely phenomenological, it is the equation associated with the parameter we use as clock. The framework itself treats all one-parameter subgroups on an equal footing.

\subsection{5.3. Time Neutrality and Background Independence}
The above derivation embodies the two structural principles,i.e., time neutrality and background independence that motivated the TRF from the outset. 

In standard quantum mechanics, time is an external parameter, a classical background against which all dynamics unfolds. The Schr\"odinger equation simply takes this time as given, without explaining what it is or why it is privileged. In the TRF, there is no such external time. What we call ``time" is nothing more than the coordinate along a particular one-parameter family of automorphisms of 
$G$—the one that happens to correlate with the ticking of laboratory clocks. Other one-parameter families correspond to other physical parameters—spatial displacement, rotation angle, gauge phase—,and each is governed by exactly the same equation of motion. No parameter is singled out as fundamental. 

Besides, there are no spacetime manifold, no background metric, no external time parameter, and no pre-existing Hilbert space appear anywhere in the derivation. The group $G$ is the totality of physical operations available to the experimentalist, and the automorphisms of $G$ encode all possible changes of physical circumstance. The general equation of motion Eq. (14) governs how the state changes under any such automorphism. When the experimentalist chooses a particular subgroup as clock, an effective Schr\"odinger equation emerges. In quantum cosmology, where no external clock exists, the abstract equation of motion Eq. (14) or Eq. (18) remains well-defined \cite{Page}. The problem of time in quantum gravity thus finds possible a natural resolution within the framework.

\section{6. Noncommutativity from Group Structure}
\label{sec6}
In the TRF, noncommutativity follows directly from the structure of the symmetry group $G$. When the group of physical operations is non-abelian, the Stone generators of its one-parameter subgroups necessarily fail to commute, and their commutation relations are completely determined by the Lie algebra of $G$. The derivation uses only a basic fact from Lie theory and the GNS construction.

Let $G$ be the symmetry group of the system, and let $\mathfrak{g}$ be its Lie algebra. The GNS construction provides a Hilbert space $\mathcal{H}_{\chi}$ and a unitary representation $U_{\chi}$. From any unitary representation of a Lie group, one obtains a representation of its Lie algebra, which is called derived representation defined as
\begin{equation}
dU_{\chi}(K)\equiv\frac{d}{ds}U_{\chi}(e^{iKs})\bigg |_{s=0}, \qquad \forall K\in\mathfrak{g}.
\end{equation}
Because each $U_{\chi}(e^{Ks})$ is unitary, its derivative $dU_{\chi}(K)$ is anti‑Hermitian $[dU_{\chi}(K)]^{\dagger}=-dU_{\chi}(K)$.
A fundamental theorem of Lie theory states that $dU_{\chi}$ is a Lie algebra homomorphism that preserves the Lie bracket \cite{Wigner}. That is, for $\forall A, B\in\mathfrak{g}$, we have
\begin{equation}
\left[dU_{\chi}(A), dU_{\chi}(B)\right]=dU_{\chi}\left(\left[A, B\right]  \right).
\end{equation}
The above equation is the mathematical statement of noncommutativity at the level of the abstract group. It tells us that if two generators fail to commute in the Lie algebra, their images as operators in the Hilbert space also fail to commute. The algebraic structure of quantum observables is thus entirely inherited from the Lie algebra of $G$. For any $K\in\mathfrak{g}$, the anti-Hermitian operator $dU_{\chi}(K)$ can be converted to a Hermitian operator by multiplying by $i$, i.e., $\hat{K}\equiv idU_{\chi}(K)$. This leads to the familiar unitary representation $U_{\chi}(e^{Ks})=e^{dU_{\chi}(K)s}=e^{-i\hat{K}s}$. We now translate the mathematical commutator into physical language.
\begin{equation}
\begin{aligned}
[\hat{A}, \hat{B}]&= [idU_{\chi}(A), idU_{\chi}(B)]=-[dU_{\chi}(A), dU_{\chi}(B)]=-dU_{\chi}([A, B]).
\end{aligned}
\end{equation}
Define the physical observable corresponding to the Lie bracket $[A, B]$ as $\widehat{[A, B]}=idU_{\chi}([A, B])$, we obtain the physical commutation theorem
\begin{equation}
\boxed{[\hat{A}, \hat{B}]=i\widehat{[A, B]}},
\end{equation}
which states that the commutator of any two Hermitian observables is $i$ times the observable that represents their Lie bracket. 

We now specialize to the case of a non‑relativistic particle. The symmetry group is the Heisenberg-Weyl group $H_{s}$. Its Lie algebra $\mathfrak{h}_{3}$is spanned by three elements $Q, P, Z$ with $[Q, P]=Z, [Q, Z]=[P, Z]=0$. Physically, $P$ generates spatial translations, $Q$ generates Galilean boosts, and $Z$ is a central element generating an overall phase. In an irreducible unitary representation, the anti-Hermitian operator $dU_{\chi}(Z)$ corresponding to the central element must, by Schur's lemma, act as a multiple of the identity. Let $dU_{\chi}(Z)=-i\hbar I$, where $\hbar$ is the same empirical calibration constant discussed above. We thus obtain the canonical commutation relation
\begin{equation}
[\hat{Q}, \hat{P}]=i\widehat{[Q, P]}=i\hat{Z}=-dU_{\chi}(Z)=i\hbar I.
\end{equation}
For rotational symmetry, the group is $SO(3)$ or its cover $SU(2)$, the real Lie algebra satisfies $[J_{i}, J_{j}]=\epsilon_{ijk}J_{k}$. The direct application of Eq. (26) gives $[\hat{J}_{i}, \hat{J}_{j}]=i\epsilon_{ijk}\hat{J}_{k}$. Physically, the angular momentum operators must carry dimensions of action when expressed in laboratory units. The calibration procedure fixes this scale factor uniquely as $\hbar$ such that $\hat{J}^{p}=\hbar\hat{J}$. We thus obtain the angular momentum commutation relation
\begin{equation}
[\hat{J}_{i}^{p}, \hat{J}_{j}^{p}]=i\hbar\epsilon_{ijk}\hat{J}_{k}^{p}.
\end{equation}

\section{7. The Path Integral as Trotter Limits of the Characteristic Function}
\label{sec52}

In standard quantum mechanics, the Feynman path integral is introduced as an independent formulation of the theory \cite{Feynman}. In the TRF, however, the Feynman path integral can be derived as the continuum limit of the characteristic function itself, via the Trotter product formula applied to the group $G$ \cite{Trotter}. This derivation reveals that the path integral is the natural analytic continuation of the response catalog, which expresses the system's response to a continuously parameterized family of transformations as an integral over the group manifold. We proceed in two stages. First, we derive a general path integral formula on the group $G$ for any one-parameter family of automorphisms. We then specialize to time evolution and recover the Feynman path integral in configuration space. Finally, we extend the construction to quantum field theory and derive the path integral representation of correlation functions.

\subsection{7.1. The General Path Integral on the Group Manifold}
Let $G$ be the symmetry group and $|\Omega\rangle_{\chi}$ the GNS reference vector. For each $g\in G$, we can define the vector $|g\rangle\equiv U_{\chi}(g)|\Omega\rangle_{\chi}$. Physically, $|g\rangle$ is obtained by applying the operation $g$ to the reference configuration of system. Because $|\Omega\rangle_{\chi}$ is cyclic, the set $\{|g\rangle: g\in G \}$ spans a dense subspace of $\mathcal{H}_{\chi}$. For any locally compact group $G$, the left Haar measure $dg$ provides a resolution of identity 
\begin{equation}
\int_{G}|g\rangle\langle g|dg=I,
\end{equation}
which is the natural completeness relation in the TRF. Now consider a one‑parameter subgroup $\{g(s)=e^{-iKs} \}_{s\in\mathbb{R}}\subset G$ with $K$ is the generator of subgroup. In the GNS representation, the characteristic function $\chi$ at a specific parameter value $S$ is given as
\begin{equation}
\chi(g(S))=_{\chi}\langle\Omega|U_{\chi}(g(S))|\Omega\rangle_{\chi}=_{\chi}\langle\Omega|e^{-i\hat{K}S}|\Omega\rangle_{\chi}.
\end{equation}
We now derive a path integral representation of $\chi(g(S))$ that is valid for any one‑parameter subgroup, without interpreting $s$ as time.

Partition the interval $[0, S]$ into $N$ equal steps of length $\epsilon=S/N$. Since $\{g(s)\}$ is a one‑parameter subgroup, $g(S)=g(\epsilon)^{N}$, and therefore the characteristic function becomes
\begin{equation}
\chi(g(S))=_{\chi}\langle\Omega|e^{-i\hat{K}S}|\Omega\rangle_{\chi}=_{\chi}\langle\Omega|(e^{-i\hat{K}\epsilon})^{N}|\Omega\rangle_{\chi}.
\end{equation}
Between each pair of factors, insert the identity Eq. (29). Denoting the intermediate group elements by $g_{1}, g_{2}, ..., g_{N-1}\in G$, with the boundary identifications $g(0)=e$ (the identity element) and $g_{N}=g(S)$, we obtain
\begin{equation}
\chi(g(S))=\int_{G}dg_{1}dg_{2}\cdots dg_{N-1}\prod_{m=1}^N\langle g_{m}|e^{-i\hat{K}\epsilon}|g_{m-1}\rangle,
\end{equation}
with the boundary vectors are $|g_{0}\rangle=|\Omega\rangle_{\chi}$ and $\langle g_{N}|=\langle\Omega|U_{\chi}(g(S))^{\dagger}$. Suppose that $N$ is sufficient large such that $\epsilon$ is small enough, we consider to compute a single factor $\langle g_{m}|e^{-i\hat{K}\epsilon}|g_{m-1}\rangle$ to first order in $\epsilon$. Define first the infinitesimally rotated generator as
\begin{equation}
\hat{K}_{g_{m-1}}\equiv U_{\chi}(g_{m-1}^{-1})\hat{K}U_{\chi}(g_{m-1}),
\end{equation}
then we have
\begin{equation}
\langle g_{m}|e^{-i\hat{K}\epsilon}|g_{m-1}\rangle=_{\chi}\langle\Omega|U_{\chi}(g_{m}^{-1})e^{-i\hat{K}\epsilon}U_{\chi}(g_{m-1})|\Omega\rangle_{\chi}=_{\chi}\langle\Omega|U_{\chi}(g_{m}^{-1}g_{m-1})e^{-i\hat{K}_{g_{m-1}}\epsilon}|\Omega\rangle_{\chi},
\end{equation}
where $e^{-i\hat{K}\epsilon}U_{\chi}(g_{m-1})=U_{\chi}(g_{m-1})e^{-i\hat{K}_{g_{m-1}}\epsilon}$ is used. For small enough $\epsilon\rightarrow 0$, $e^{i\hat{K}_{g_{m-1}}\epsilon}=I-i\hat{K}_{g_{m-1}}\epsilon+o(\epsilon^{2})$, 
the above equation becomes
\begin{equation}
\begin{aligned}
\langle g_{m}|e^{-i\hat{K}\epsilon}|g_{m-1}\rangle &=_{\chi}\langle\Omega|U_{\chi}(g_{m}^{-1}g_{m-1})|\Omega\rangle_{\chi}-i\epsilon_{\chi}\langle\Omega|U_{\chi}(g_{m}^{-1}g_{m-1})\hat{K}_{g_{m-1}}|\Omega\rangle_{\chi}+o(\epsilon^{2}) \\
&=\chi(g_{m}^{-1}g_{m-1})-i\epsilon\langle g_{m}|\hat{K}|g_{m-1}\rangle + o(\epsilon^{2}) \\
&=\chi(g_{m}^{-1}g_{m-1})\mathrm{exp}\left[-i\epsilon\frac{\langle g_{m}|\hat{K}|g_{m-1}\rangle}{\chi(g_{m}^{-1}g_{m-1})}\right]+ o(\epsilon^{2}) \\
&=\langle g_{m}|g_{m-1}\rangle\mathrm{exp}\left[-i\epsilon\frac{\langle g_{m}|\hat{K}|g_{m-1}\rangle}{\langle g_{m}|g_{m-1}\rangle}\right]+ o(\epsilon^{2})
\end{aligned}
\end{equation}
In the continuum limit $N\rightarrow\infty, \epsilon\rightarrow 0$ with $N\epsilon=S$ fixed, the intermediate group elements define a continuous path $g(s)$ in $G$ with $g(0)=e$ and $g(S)=g_{S}$, and we obtain 
\begin{equation}
\langle g_{m}|g_{m-1}\rangle =\langle g(s+\epsilon)|g(s)\rangle=\langle g(s)|g(s)\rangle+\epsilon\langle\partial_{s}g|g\rangle+o(\epsilon^{2}).
\end{equation}
Take the logarithm of $\langle g_{m}|g_{m-1}\rangle$ gives $\mathrm{ln}\langle g_{m}|g_{m-1}\rangle=\epsilon\langle\partial_{s}g|g\rangle+o(\epsilon^{2})$. Summing over all $m=1, \cdots, N$ and taking the continuum limit $N\rightarrow\infty$ gives
\begin{equation}
\begin{aligned}
&\lim_{N\rightarrow\infty}\sum_{m=1}^{N}\mathrm{ln}\langle g_{m}|g_{m-1}\rangle =\int_{0}^{S}ds\langle\partial_{s}g|g\rangle=-\int_{0}^{S}ds_{\chi}\langle\Omega|U_{\chi}(g^{-1})\frac{d}{ds}U_{\chi}(g)|\Omega\rangle_{\chi}, \\
&\lim_{\epsilon\rightarrow 0}\frac{\langle g_{m}|\hat{K}|g_{m-1}\rangle}{\langle g_{m}|g_{m-1}\rangle}=\langle g(s)|\hat{K}|g(s)\rangle\equiv k(g(s)).
\end{aligned}
\end{equation}
The first term of Eq. (37) actually defines the {\bf Berry term}, a geometric phase that depends only on the path $g(s)$ traced in the group manifold, not on the parametrization. We can therefore define the so-called {\bf action functional on the group manifold}:
\begin{equation}
\boxed{\boldsymbol{S}[g]=\int_{0}^{S}ds[i\hbar\langle g(s)|\partial_{s}|g(s)\rangle-k(g(s))]}, 
\end{equation}
and thus rewritten Eq. (26) as
\begin{equation}
\boxed{\chi(g(S))=\int_{g(0)=e}^{g(S)=g_{S}}\mathcal{D}g(s)\mathrm{exp}\left(\frac{i}{\hbar}\boldsymbol{S}[g]\right)},
\end{equation}
where the functional measure $\mathcal{D}g(s)$ is the continuum limit of the product of Haar measures $\Pi_{m=1}^{N-1}dg_{m}$, together with the normalisation factors from the $\chi$ prefactors. Eq. (39) is the most general path integral in the TRF. It holds for any one‑parameter subgroup $\{g_{s}\in G\}$, labelled by an arbitrary parameter $s$.

\subsection{7.2. Recovering the Feynman Path Integral}
We now show how the familiar Feynman path integral emerges as a special case when three physically motivated choices are made: (1) The symmetry group $G$ is the Heisenberg–Weyl group; (2) The one‑parameter subgroup is identified with time evolution such that $\hat{K}=\hat{H}$; (3) The reference vector is chosen to yield canonical coherent states.

For a non‑relativistic particle, the fundamental operations are translations in position and momentum, which is described by the Heisenberg–Weyl group $H_{3}$. The group elements of $H_{s}$ can be parameterized as $U_{\chi}(g)=e^{-i\hat{Q}p}e^{i\hat{P}q}e^{i\phi}$ in the GNS representation. The reference vector $|\Omega\rangle_{\chi}$ is chosen to be the ground state of a harmonic oscillator with a specific frequency $w_{0}$. Applying a general group element to this reference state produces the canonical coherent states
\begin{equation}
|g\rangle=U_{\chi}(g)|\Omega\rangle_{\chi}=e^{-i\hat{Q}p}e^{i\hat{P}q}e^{i\phi}|\Omega\rangle_{\chi}\equiv|q, p\rangle
\end{equation}
The resolution of identity for these coherent states $|q, p\rangle$ is
\begin{equation}
\int\frac{dqdp}{2\pi\hbar}|q,p\rangle\langle q,p|=I.
\end{equation}
Calculating the Berry term in the action functional $\boldsymbol{S}[g]$ gives
\begin{equation}
i\hbar\langle q,p|\partial_{t}|q,p\rangle=\frac{1}{2}(p\dot q-q\dot p)=p\dot q-\frac{1}{2}\frac{d}{dt}(pq).
\end{equation}
The second term is calculated as 
\begin{equation}
\langle q,p|\hat{H}|q,p\rangle=\langle q,p|\frac{\hat{P}^{2}}{2m}+V(\hat{Q})|q,p\rangle=\frac{p^{2}}{2m}+V(q)=H(q,p),
\end{equation}
where we have omitted constant term and $o(\hbar)$ term.
Note that the total derivative inside the action only contributes a boundary phase factor and leaves the equations of motion as well as the kernel of the path integral unchanged. We thus obtain the action in phase space
\begin{equation}
\boldsymbol{S}[q,p]=\int_{0}^{T}dt[p\dot q-H(q,p)]=\int_{0}^{T}dt\mathcal{L}(q,p).
\end{equation}
The characteristic function $\chi(g_{T})$ becomes the phase-space path integral:
\begin{equation}
\chi(g_{T})=\int\mathcal{D}q\mathcal{D}p e^{\frac{i}{\hbar}\int_{0}^{T}dt\mathcal{L}(q,p)},
\end{equation}
where the functional measure is the continuum limit of the product of Haar measures
\begin{equation}
\int\mathcal{D}q\mathcal{D}p=\lim_{N\rightarrow\infty}\int\prod_{m=1}^{N-1}\frac{dq_{k}dp_{k}}{2\pi\hbar}.
\end{equation}

The phase-space path integral contains an integration over the momentum variables $p(t)$ at each time slice. Because the exponent is quadratic in $p$, these integrals are Gaussian and can be performed exactly. This is the crucial step that transforms the phase-space path integral into the configuration-space Feynman path integral.
At each time slice $t_{k}=k\epsilon$, the relevant factor in the discrete form of Eq. (45) is
\begin{equation}
\int_{-\infty}^{\infty}\frac{dp_{k}}{2\pi\hbar}\mathrm{exp}\left[\frac{i}{\hbar}p_{k}\Delta q_{k}-\frac{i\epsilon}{2m\hbar}p_{k}^{2}\right]=\sqrt{\frac{m}{2\pi i\hbar\epsilon}}\mathrm{exp}\left[\frac{i}{\hbar}\epsilon\left(\frac{\Delta q_{k}}{\epsilon}\right)^{2} \right]\times e^{-\frac{i}{\hbar}\epsilon V(q_{k-1})},
\end{equation}
where $\Delta q_{k}\equiv q_{k}-q_{k-1}$ and Fresnel integral formula is used. Now take the product over all 
$N$ time slices and the continuum limit $N\rightarrow\infty, \epsilon\rightarrow 0$ with $N\epsilon=T$ fixed, we thus obtain the Feynman path integral in configuration space
\begin{equation}
\chi(g_{T})=\int dq_{f}dq_{i}\psi_{0}(q_{f})^{*}\psi_{0}(q_{i})\int_{q(0)=q_{i}}^{q(T)=q_{f}}\mathcal{D}q(t)\mathrm{exp}\left(\frac{i}{\hbar}\int_{0}^{T}dt\left[\frac{1}{2}m\dot q^{2}-V(q)\right]\right),
\end{equation}
where $\psi_{0}(q_{i})=\langle q_{i}|\Omega\rangle_{\chi}$.

\subsection{7.3. The Path Integral in Quantum Field Theory}
Here we derive the functional path integral for a generic quantum field theory, emphasizing that every ingredient of quantum field theory emerges from the group structure and the positive-definiteness condition, without any recourse to canonical quantization \cite{Haag}.
For a field theory in $4$-dimensional Minkowski spacetime, the operations fall into two classes: (1) at each spacetime point $x$, the experimenter can shift the field value by a real parameter $\lambda$, which generates the infinite‑dimensional group $G_{int}$ of local field transformations; (2) the experimenter can translate, rotate, and boost the entire physical system, and these operations form the Poincar\'e group $G_{\mathcal{P}}$ in flat spacetime. The two classes of operations are intertwined. A local field translation must be performed at a specific spacetime point, and the labelling of points is provided by the Poincar\'e group. Mathematically, the full symmetry group is the semidirect product $G_{QFT}=G_{int} \rtimes G_{\mathcal{P}}$.

A local field translation by an amount $\lambda$ at the spacetime point $x$ corresponds to a one‑parameter subgroup $\{ g_{\lambda}(x)\}\subset G_{int}$, which we have $U_{\chi}(g_{\lambda}(x))=e^{-i\lambda\hat{\phi}(x)}$ in the GNS representation. A general group element of $G_{int}$ can therefore be written as
\begin{equation}
U_{\chi}(g(J))=\mathrm{exp}\left(i\int d^{4}x J(x)\hat{\phi}(x)  \right),
\end{equation}
where $J(x)$ is an arbitrary real-valued function playing the role of a continuous source. The position label $x$ on $\phi(x)$ is inherited from the Poincar\'e factor in the semidirect product; it guarantees that field operators at different points are related by spacetime translations in the standard way. In the TRF, field operators $\hat{\phi}(x)$ emerge naturally as the generators of the representation of an infinite-dimensional local transformation group with no need of second quantization. 

For the physical choice where $\chi$ is Poincar\'e-invariant (i.e., $\chi(g)=1$ for all $g\in G_{\mathcal{P}}$) and simultaneously minimizes the energy under local translations, the resulting cyclic reference vector $|\Omega\rangle_{\chi}$ obtained from the GNS construction is identified with the vacuum vector $|0\rangle$. This vector is distinguished as the unique cyclic vector that is annihilated by all generators of Poincar\'e transformations as well as by the positive-frequency part of the local field operators (i.e., $U_{\chi}(g)|\Omega\rangle_{\chi}=|\Omega\rangle_{\chi}$ for all $g\in G_{\mathcal{P}}$, and $\hat{H}|\Omega\rangle_{\chi}=0$), thereby reproducing the standard vacuum properties of quantum field theory without any recourse to second quantization. For the vacuum vector $|0\rangle_{\chi}$, the evaluation of  characteristic function $\chi$ on the local field translation group $G_{int}$ immediately yields the {\bf generating functional} of quantum field theory:
\begin{equation}
Z(J)\equiv\chi_{0}(g(J))=_{\chi}\langle \Omega|U_{\chi}(g(J))|\Omega\rangle_{\chi}=\langle 0|\mathcal{T}\mathrm{exp}\left(i\int d^{4}x J(x)\hat{\phi}(x)\right)|0\rangle.
\end{equation}
The above equation gives a direct operational definition of $Z(J)$. It is the response catalog of the vacuum state $\chi_{0}$ to the continuous source operation $g(J)$. The time‑ordering symbol $\mathcal{T}$ appears because the source acts over a finite time interval, and field operators at different times do not commute; it is the natural outcome of decomposing a single group element into time-ordered products of infinitesimal operations. All $n$-point correlation functions can be obtained from $Z(J)$ by functional differentiation:
\begin{equation}
G^{n}\{x_{1},\cdots, x_{n} \}=\langle 0|\mathcal{T}\{\hat{\phi}(x_{1})\cdots\hat{\phi}(x_{n})\}|0\rangle=(-i)^{n}\frac{\delta^{n}Z(J)}{\delta J(x_{1})\cdots\delta J(x_{n})}\big|_{J=0}.
\end{equation}
The correlation functions are operationally the expansion coefficients of the characteristic function $\chi_{0}$ in powers of the local transformation parameters, quantifying how the joint response to multiple simultaneous operations deviates from the product of individual responses.

We now derive the functional path-integral representation of $Z(J)$ using only the group structure and the Trotter product formula. First we note that the source $J(x)$ alone does not generate time evolution. Time evolution is driven by the Hamiltonian $\hat{H}$, which is the Stone generator of the time‑translation subgroup of the Poincar\'e group. To obtain the path integral we isolate the time‑evolution operator. Split the time interval into $N$ steps of length $\epsilon=T/N$ (with $T\rightarrow\infty$ at the end). The time‑ordered product can be written as an alternating product of infinitesimal time evolutions and source operations:
\begin{equation}
U_{\chi}(g(J))=\lim_{N\rightarrow\infty}e^{-i\hat{H}\epsilon}e^{i\epsilon\int d^{3}xJ(t_{N},\vec{x})\hat{\phi}(\vec{x})}\cdots e^{-i\hat{H}\epsilon}e^{i\epsilon\int d^{3}xJ(t_{1},\vec{x})\hat{\phi}(\vec{x})},
\end{equation}
where the order of the factors reproduces the time ordering of the operators. Let $|\phi\rangle$ be a simultaneous eigenvector of the field operators $\hat{\phi}(\vec{x})$ at a fixed time, i.e., $\hat{\phi}(\vec{x})|\phi\rangle=\phi(\vec{x})|\phi\rangle$, we have the identity resolution $I=\int\mathcal{D}\phi|\phi\rangle\langle\phi|$. Insert this resolution of the identity between every factor in Eq. (52), and denote the field configuration at time $t_{k}$ by $\phi_{k}(\vec{x})$, we obtain
\begin{equation}
Z(J)=\langle 0|U_{\chi}(g(J))|0\rangle=\lim_{N\rightarrow\infty}\int\prod_{k=1}^{N}\mathcal{D}\phi_{k}\prod_{k=1}^{N}\langle\phi_{k}|e^{-i\hat{H}\epsilon}e^{i\epsilon\int d^{3}x J(t_{k},\vec{x})\hat{\phi}(\vec{x})}|\phi_{k-1}\rangle,
\end{equation}
with the boundary condition $|\phi_{0}\rangle=|\phi_{N}\rangle=|0\rangle$. Since $|\phi_{k-1}\rangle$ is an eigenvector of $\hat{\phi}$, the exponentiated source operator acts as a phase factor to first order in $\epsilon$:
\begin{equation}
e^{i\epsilon\int d^{3}x J(t_{k},\vec{x})\hat{\phi}(\vec{x})}|\phi_{k-1}\rangle=e^{i\epsilon\int d^{3}x J(t_{k},\vec{x})\phi_{k-1}(\vec{x})}|\phi_{k-1}\rangle+o(\epsilon^{2}).
\end{equation}
Thus the source contributes a factor $e^{i\epsilon\int J\phi_{k-1}}$ and does not affect the kinetic part. For a free scalar field the Hamiltonian is
\begin{equation}
\hat{H}=\int d^{3}x\frac{1}{2}\left(\hat{\pi}^{2}+(\nabla\hat{\phi})^{2}+m^{2}\hat{\phi}^{2} \right).
\end{equation}
Trotterise the exponential as:
\begin{equation}
e^{-i\hat{H}\epsilon}=e^{-i\frac{\epsilon}{2}\int\hat{\pi}^{2}}e^{-i\frac{\epsilon}{2}\int V(\hat{\phi})^{2}}+o(\epsilon^{2}),
\end{equation}
with $V(\hat{\phi})=[(\nabla\hat{\phi})^{2}+m^{2}\hat{\phi}^{2}]/2$. Let $|\pi\rangle$ be eigenvector of $\hat{\pi}$, we have $\hat{\pi}(\vec{x})|\pi\rangle=\pi(\vec{x})|\pi\rangle, I=\int\mathcal{D}\pi|\pi\rangle\langle\pi|$ and $\langle\phi|\pi\rangle=e^{i\int d^{3}x\pi\phi}$. Inserting the identity between the kinetic and potential factors and omit $o(\epsilon^{2})$ term gives
\begin{equation}
\begin{aligned}
\langle\phi_{k}|e^{-i\hat{H}\epsilon}|\phi_{k-1}\rangle &=\int\mathcal{D}\pi_{k}\langle\phi_{k}|\pi_{k}\rangle\langle\pi_{k}|e^{-i\frac{\epsilon}{2}\int\hat{\pi}^{2}}e^{-i\frac{\epsilon}{2}\int V(\hat{\phi})^{2}}|\phi_{k-1}\rangle \\
&=\int\mathcal{D}\pi_{k}e^{i\pi_{k}(\phi_{k}-\phi_{k-1})-i\epsilon\pi_{k}^{2}/2}e^{-i\epsilon\int d^{3}x V(\phi_{k})} \\
&=\mathcal{N}_{\epsilon}e^{\frac{i}{2\epsilon}\int d^{3}x(\phi_{k}-\phi_{k-1})^{2}-i\epsilon V(\phi_{k})    },
\end{aligned}
\end{equation}
where Fresnel formula is used to perform the Gaussian integral over $\pi_{k}$ and $\mathcal{N}_{\epsilon}$ is a field‑independent normalization factor. Multiplying the above factors for all time slices and including the source phases $e^{i\epsilon\int J\phi_{k-1}}$, the total exponent becomes
\begin{equation}
\begin{aligned}
M_{k}&=\prod_{k=1}^{N}\langle\phi_{k}|e^{-i\hat{H}\epsilon}e^{i\epsilon\int d^{3}x J(t_{k},\vec{x})\hat{\phi}(\vec{x})}|\phi_{k-1}\rangle \\
&=\mathcal{N}_{\epsilon}^{N}\mathrm{exp}\left[\sum_{k=1}^{N}\left(\frac{i}{2\epsilon}\int d^{3}x(\phi_{k}-\phi_{k-1})^{2}-i\epsilon V(\phi_{k-1})+\int d^{3}x J\phi_{k-1}     \right)    \right]
\end{aligned}
\end{equation}
As $N\rightarrow\infty, \epsilon\rightarrow 0$
\begin{equation}
\frac{\phi_{k}-\phi_{k-1}}{\epsilon}\rightarrow\partial_{t}\phi, \qquad \epsilon\sum_{k}\rightarrow\int dt.
\end{equation}
Putting all elements together, we thus obtain the functional path-integral formulation
\begin{equation}
\begin{aligned}
Z(J) &= \lim_{N\rightarrow\infty}\int\prod_{k=1}^{N}\mathcal{D}\phi_{k} M_{k} \\
&=\lim_{N\rightarrow\infty}\mathcal{N}_{\epsilon}^{N}\int\prod_{k=1}^{N}\mathcal{D}\phi_{k}\mathrm{exp}\left(\frac{i}{\hbar}\int d^{4}x\left[\frac{1}{2}(\partial_{u}\phi)^{2}-\frac{1}{2}m^{2}\phi^{2}+J(x)\phi(x) \right] \right) \\
&=\int\mathcal{D}\phi\mathrm{exp}\left(\frac{i}{\hbar}S[\phi]+\int d^{4}x J(x)\phi(x) \right).
\end{aligned}
\end{equation}

\section{8. Product Order Positivity: A Possible New Physical Constraint}
\label{sec6}
The TRF has so far derived the entire standard apparatus of quantum mechanics from a single postulate. If the framework did only this, it would be an elegant reformulation. However, the 
operational logic of the framework naturally supports a structure that standard quantum mechanics does not directly encode, i.e., the superposition of operation sequences with different product orders.

\subsection{8.1. The Product Order Positivity}
The characteristic function $\chi$ is defined on the group $G$. Through linear extension it becomes a positive functional on the group algebra $C[G]$ that consists of all formal finite linear combinations of group elements with complex coefficients. For any $a=\sum_{i}c_{i}g_{i}\in C[G]$, which physically describes the superposition operation, the response to such a operation is given by $\chi(a)=\sum_{i}c_{i}\chi(g_{i})$. The positive-definiteness of $\chi$ on group $G$ guarantee the positive of $\chi$ on group algebra $C[G]$, which gives $\chi(a^{*}a)\ge 0$ with $a^{*}=\sum_{i}c_{i}^{*}g_{i}^{-1}$ is the involution on the $C[G]$. This condition ensures that every superposition of operations yields a non-negative probability for the system to return to its reference configuration. 
Since $gh$ and $hg$ are also elements of group $G$, the group algebra $C[G]$ naturally support the superposition of operation sequences $\alpha(gh)+\beta(hg)$ with $g,h$ do not commute. It is a legitimate operation in the TRF because the experimenter can realize it by coupling the system to a quantum controller that determines the order in superposition, which is precisely the structure exploited in the quantum switch. 

An experimenter may choose to restrict attention to operations performed in a fixed operation order. Suppose that only responses to sequences in which operation $g$ precedes $h$ are recorded. The set of such operations is denoted as $S_{g<h}=\{hg|\forall g, h\in G, gh\ne hg\}$. The set $S_{g<h}$ is not a subgroup of $G$ as it is clear that $(hg)^{-1}=g^{-1}h^{-1}$ is not in the set. However, the experimenter can still organize data by labeling each sequence by the ordered pair $(g,h)$. The set of such pairs forms the direct product group $H_{g<h}=G\times G$ with multiplication $(g_{1},h_{1})\cdot(g_{2},h_{2})=(g_{1}g_{2},h_{1}h_{2})$. The map
\begin{equation}
\iota_{g<h}: H_{g<h}\rightarrow G, \iota_{g<h}(g, h)=hg
\end{equation}
embeds the ordered sector into the full group. Crucially, $\iota_{g<h}$ is not a group homomorphism unless all elements of $G$ commute, which implies that
\begin{equation}
\iota_{g<h}((g_{1},h_{1})(g_{2},h_{2}))\ne\iota_{g<h}(g_{1},h_{1})\iota(g_{2},h_{2}).
\end{equation}
The ordered-sector characteristic function is the pullback of $\chi$ by the embedding $\chi_{g<h}(g,h)\equiv\chi(\iota_{g<h}(g,h))=\chi(hg)$, which encodes all responses the experimenter can record when restricts operations to the order $g$ first and $h$ second. The operational logic of the TRF demands that any self-consistent set of response data define a valid quantum state. An experimenter confined to the ordered sector $H_{g<h}$ should therefore be able to describe data by a positive-definite function on that sector. This yields the {\bf product order positivity condition: $\chi_{g<h}$ is positive-definite on $H_{g<h}$}. Explicitly, for any finite collection $\{(g_{i},h_{i})\subset H_{g<h} \}$ and any complex coefficients $\{c_{i}\}$
\begin{equation}
\sum_{i,j}c_{i}^{*}c_{j}\chi_{g<h}((g_{1},h_{i})^{-1}(g_{j},h_{j}))=\sum_{i,j}c_{i}^{*}c_{j}\chi(h_{i}^{-1}h_{j}g_{i}^{-1}g_{j})\ge 0.
\end{equation}
The same condition must hold for the opposite order $h<g$, with characteristic function $\chi_{h<g}(g,h)=gh$. {\bf The product order positivity is not a logical consequence of the global positive-definiteness postulate when $G$ is non‑abelian}. This is due to the fact that the embedding $\iota_{g<h}$ is not a $*$-homomorphism with respect to the natural involutions $\iota_{g<h}((g,h)^{*})=\iota_{g<h}(g^{-1},h^{-1})=h^{-1}g^{-1}\ne g^{-1}h^{-1}=\iota_{g<h}(g,h)^{*}$.

\subsection{8.2. Indefinite Causal Order and the Quantum Switch}

The physical significance of product order positivity becomes sharpest in the setting of indefinite causal order (ICO) \cite{ICO1, ICO2, ICO3}. The quantum switch \cite{switch, S2, S3, S4}, the paradigmatic ICO protocol, provides a concrete experimental context in which product order positivity can be tested. Consider a physical system with symmetry group $G$, in which $g,h\in G$ be two non-commuting operations. In the quantum switch situation, an experimenter arranges a coherent superposition of the two product orders $w=\alpha\cdot(gh)+\beta\cdot(hg)\in C[G]$ with $|\alpha|^{2}+|\beta|^{2}=1$. This operation is a legitimate element of the group algebra, and its realization does not require the introduction of an external control qubit, although in practice such a qubit provides a convenient implementation. What matters for the TRF is that the experimenter can, by coupling the system to an appropriate controller, engineer a process whose net effect on the system is given by $w$. 

The global characteristic function $\chi$ on $G$ is positive-definite by the core postulate. The response to the superposition operation is, by linear extension, $\chi(w)=\alpha\chi(gh)+\beta\chi(hg)$. Now, the experimenter may ask a more refined question: what happens if the observations are restricted to operations performed in a fixed order? For instance, the experimenter may choose to record only those events in which the operation $g$ is applied before 
$h$. In the language of the characteristic function, this corresponds to extracting the ordered‑sector function defined as $\chi_{g<h}(g,h)=\chi(hg)$. This function is obtained directly from the global $\chi$ by evaluating it on the specific group elements $hg$. The experimenter simply prepares the system, applies the composite operation $hg$ with $g$ first, and records the interference response. She does this for many different pairs $(g,h)$, thereby building the catalog $\chi_{g<h}$. Product order positivity demands that this catalog be a positive-definite function on the ordered group $H_{g<h}$. For the quantum switch, this means that even though the full process involves a superposition of orders, the data restricted to a definite order must form a valid quantum state in its own right. If they do not, then an experimenter who only has access to the fixed-order sector would observe data inconsistent with the axioms of the TRF. The experimenter would be forced to conclude that no physical system exists, contradicting the fact that the unrestricted experimenter verifies a well‑defined global state via the same operational criteria.
The product order positivity thus constrains the admissible forms of $\chi$ on the full group $G$. For a given superposition coefficient $\alpha, \beta$, the global characteristic function must be such that both $\chi_{g<h}$ and $\chi_{h<g}$ are positive-definite. This is a non‑trivial restriction when $G$ is non-abelian, precisely because the algebraic structure of the ordered sector differs from that of the ambient group.

In practice, testing product order positivity condition with a quantum switch involves considering the characteristic function $\chi$ for many different pairs $(g,h)$ and then constructing the matrix $M_{ij}=\chi_{g<h}(g_{i}^{-1}g_{j},h_{i}^{-1}h_{j})=\chi(h_{i}^{-1}h_{j}g_{i}^{-1}g_{j})$, where $M_{ij}$ can actually be direct measured in the experiment with composite operations $h_{i}^{-1}h_{j}g_{i}^{-1}g_{j}$ are applied. The positive semi-definiteness of this matrix for arbitrary finite sets of pairs constitutes a direct experimental check of product order positivity. Existing quantum switch platforms in photonic system \cite{P1, P2, P3} have not yet performed these specific measurements, but the protocol is feasible with current technology.
The significance of the quantum switch for product order positivity is not that it introduces a control qubit, but that it provides a physical realization of the group-algebra element $w$. The TRF then demands that the ordered-sector responses extracted from this realization must themselves satisfy the axioms of the framework. 
Besides, product order positivity may constrain the violation of causal inequalities in ICO processes. The Tsirelson bound arises in the standard process matrix framework from the requirement that the process matrix be positive semi-definite \cite{ICO3}. Since product order positivity imposes additional conditions beyond global positivity, specifically the positivity of the ordered‑sector reductions, it may tighten this bound or, at minimum, provide an alternative derivation of it from purely group-algebraic principles.
Whether nature complies is an experimental question, one that transforms product order positivity from a formal condition into a falsifiable prediction.

\section{9. Discussion and Conclusion}
\label{sec6}
The TRF is not actually limited to quantum mechanics. It is a general operational logic for physical theories, grounded in a single principle that physical state is defined as response catalog to all physical transformation operations. The mathematical machinery, the GNS construction, Bochner theorem, the commutation theorem, the Trotter limit, requires only a group and a positive‑definite function. What changes, when one changes the group, is the physics. Choose the Poincar\'e group, and the framework may yield relativistic quantum field theory. Choose the gauge groups of the Standard Model, it may provide the algebraic skeleton of particle physics.
Choose the group of spatial diffeomorphisms, and it provides a background‑independent kinematic structure that may serve as a starting point for a quantum theory of gravity. The framework itself is indifferent to these choices. This universality is its deepest promise: the operational logic distilled here from quantum mechanics may, in fact, be the operational logic of any fundamental physical theory.

The present work has focused on establishing this logic for quantum mechanics. Starting from the single postulate that a quantum state is described by a positive-definite function $\chi$ on the symmetry group $G$, we have derived the entire standard apparatus of the quantum theory. The Hilbert space emerged as the GNS representation of the response catalog. The Born rule appeared as Bochner theorem applied to abelian subgroups. The Schr\"odinger equation followed from the action of group automorphisms on the characteristic function. The commutation relations of all observables were shown to be consequences of the Lie algebra of the symmetry group. The Feynman path integral, in both its quantum mechanical and quantum field-theoretic forms, was recovered as the Trotter continuum limit of the characteristic function. At no point were any of the standard axioms of quantum mechanics invoked as independent postulates, they are all theorems.

The framework also yields a concrete physical constraint that distinguishes it from both the standard formulation and earlier axiomatic reconstructions. Product order positivity, which requires that the characteristic function, when restricted to an operationally defined ordered sector, itself defines a positive-definite function on that sector. It arises from the group-algebra structure that supports superpositions of operation sequences. It is operationally natural: an experimenter confined to a fixed product order must be able to describe the measured data as a valid quantum state. It is mathematically non‑trivial: the embedding of an ordered sector into the full group is not a $*$-homomorphism when the group is non-abelian, so the condition does not reduce to global positive-definiteness. Most importantly, it is, in principle, experimentally accessible. Quantum-switch protocols can test whether the ordered-sector reductions of a globally admissible process satisfy positive-definiteness. Whether product order positivity is strictly independent of global positive-definiteness, and whether it can rule out process matrices that are mathematically valid in the standard ICO framework, are open questions that define an immediate research programme.

Much remains to be done. The mathematical status of product order positivity must be settled. Its connection to the Tsirelson bound for causal inequalities, and to the process matrix framework more generally, must be clarified. The framework's extension to quantum gravity, where the symmetry group itself may be emergent, is a direction rich with promise but largely unexplored. The classical limit, understood as the abelianisation of the symmetry group, must be connected to the concrete mechanisms of decoherence. The construction of interacting quantum field theories within the framework, and the fate of renormalization when the response catalog is taken as primitive, remain open. These are substantial challenges. But a framework that answers every question is a closed book. A framework that opens new questions is alive. The TRF reduces the axiomatic burden of quantum mechanics to a single physical condition, and from this condition extracts not only the known structure of the theory but also a previously unrecognized physical constraint. Whether this constraint survives experimental scrutiny will determine whether the framework is a reformulation or a genuine extension of quantum mechanics. The question is now in the hands of experiment.

\begin{acknowledgments}
{\bf Acknowledgments:} 
The author sincerely thanks DeepSeek V4 Pro for assistance in drafting and editing the manuscript, including language polishing, formatting of mathematical expressions, and the articulation of certain parts of the theoretical derivations. All core physical ideas, the final validation of all derivations, and full academic responsibility remain solely with the author.
This work is supported by the National Key Research and Development
Program of China (2025YFE0217200).
\end{acknowledgments}











\end{document}